\documentclass{article}

\usepackage{PRIMEarxiv}

\usepackage[utf8]{inputenc} 
\usepackage[T1]{fontenc}    
\usepackage{hyperref}       
\usepackage{url}            
\usepackage{booktabs}       
\usepackage{amsfonts}       
\usepackage{nicefrac}       
\usepackage{microtype}      
\usepackage{lipsum}
\usepackage{fancyhdr}       
\usepackage{graphicx}       
\graphicspath{{media/}}     

\usepackage{amssymb}
\usepackage{tabularx}
\usepackage{threeparttable}
\usepackage{csquotes}
\usepackage{textgreek}
\usepackage{float}
\usepackage{siunitx}
\usepackage{stmaryrd}

\usepackage{physics}
\usepackage{multirow}

\usepackage{caption}
\usepackage{subcaption}
\usepackage{xcolor}

\usepackage{booktabs}
\usepackage{array} 
\usepackage{colortbl}
\usepackage{diagbox}
\usepackage{stmaryrd}
\usepackage{empheq, etoolbox}
\title{High-fidelity Grain Growth Modeling: Leveraging Deep Learning for Fast Computations}

\author{Pungponhavoan ~Tep\thanks{Corresponding author: pungponhavoan.tep@minesparis.psl.eu} , 
Marc ~Bernacki \\
	\\
	Mines Paris, PSL University\\
 	Centre for material forming (CEMEF), UMR CNRS\\
 	 06904 Sophia Antipolis, France\\
}

\begin{document}
\maketitle

\begin{abstract}
Grain growth simulation is crucial for predicting metallic material microstructure evolution during annealing and resulting final mechanical properties, but traditional partial differential equation-based methods are computationally expensive, creating bottlenecks in materials design and manufacturing. In this work, we introduce a machine learning framework that combines a Convolutional Long Short-Term Memory networks with an Autoencoder to efficiently predict grain growth evolution. Our approach captures both spatial and temporal aspects of grain evolution while encoding high-dimensional grain structure data into a compact latent space for pattern learning, enhanced by a novel composite loss function combining Mean Squared Error, Structural Similarity Index Measurement, and Boundary Preservation to maintain structural integrity of grain boundary topology of the prediction. Results demonstrated that our machine learning approach accelerates grain growth prediction by up to \SI{89}{\times} faster, reducing computation time from \SI{10}{\minute} to approximately \SI{10}{\second} while maintaining high-fidelity predictions. The best model (S-30-30) achieving a structural similarity score of \SI{86.71}{\percent} and mean grain size error of just \SI{0.07}{\percent}. All models accurately captured grain boundary topology, morphology, and size distributions. This approach enables rapid microstructural prediction for applications where conventional simulations are prohibitively time-consuming, potentially accelerating innovation in materials science and manufacturing.
\end{abstract}



\section{Introduction} 
\label{Introduction}
In material science, understanding the evolutions of microstructure is crucial for predicting and improving the properties of materials under various external conditions, such as thermal treatment, applied stress, and other environmental factors \cite{Choudhary2022}. Grain growth is considered as one critical mechanism among various microstructural evolution mechanisms as it significantly influences both mechanical and physical properties of a material ranging from strength, toughness and corrosion resistance \cite{callister-2013}. For example, in steel, the size of grains and their distribution have a significant impact on yield strength, hardness, and fracture toughness. Finer grain structure generally leads to improved fracture toughness, whereas larger grain structure tends to enhance creep properties under high temperature conditions. This microstructure evolution mechanism typically happens during the annealing processes in which larger grains grow systematically at the expense of smaller grains to minimize the total grain boundary energy within the system, occurs through the migration of grain boundaries driven by reduction in the total grain boundary energy. In continuum mechanics, the migration of the grain boundaries is classically modeled by the following curvature flow kinetic equation \cite{BERNACKI2024101224}:

\begin{equation}
    \vec{V} = - \mu \gamma \kappa \vec{n} \quad \label{eq:grain_growth_equation}
\end{equation}

where \( \vec{V} \) is the velocity, \(\mu\) is the temperature-dependent mobility, \( \gamma \) is grain boundary energy, \(\kappa\) is the trace of the curvature tensor, and \(\vec{n}\) is the outward unit normal vector of the grain boundary. The equation suggests that for a homogeneous spatial value of the reduced mobility ($\mu\gamma$ product), grain growth is primarily driven by the curvature of grain boundaries, with higher-curvature regions moving faster than those with lower-curvature regions. The representativeness of this equation—when coupled with the conventional five-dimensional (5-D) space and models describing grain boundary energy and mobility—is increasingly debated in recent studies \cite{Chen2020,Bhattacharya2021,Xu2023,Florez2022,qiu2025}. These works ultimately emphasize that grain growth is not just curvature flow and that the equation remains a first order approximation. Nevertheless, at the polycrystalline scale, it remains a statistically excellent approximation, as extensively documented in the literature through numerous comparisons between experimental data and high-fidelity mesoscopic simulations.

Numerous computational methods have been developed to study the effects of the mechanism including probabilistic models such as Monte Carlo Potts \cite{rollett2004monte} and cellular automata \cite{janssens-2009,golab-2014}, as well as deterministic methods based on partial-differential equations (PDEs) such as phase-field methods \cite{moelans2008quantitative,krill-2002}, front-tracking/vertex approaches \cite{mora2008three,florez2020novel}, and level-set models \cite{zhao1996variational,BERNACKI2024101224,hallberg2019modeling}. While these methods are widely used and demonstrates robust capabilities in predicting the evolution, they impose a significant limitation which is the computational cost. For instance, the deterministic PDE-based models typically relies on sophisticated mathematical formulations to characterize the grain growth, requiring iterative calculations at each temporal increment to update the microstructure state.

In industrial contexts, simulation tools based on such methods are repeatedly performed and scaled to large systems in predicting the evolution of microstructure in order to optimize material properties. Despite probabilistic approaches (like Monte Carlo Potts and cellular automata) being more computationally efficient than their deterministic counterparts, these approaches typically span several hours to days or even weeks when applied to industrially relevant scales and complexities. This makes rapid iteration less practical, specifically when multiple simulations are required for parameters fine-tuning or time-sensitive material design tasks.

Machine learning (ML) introduced a paradigm shift in computational material science, offering new approaches to modeling complex physical phenomena \cite{himanen2019data}. In contrast to the conventional simulation methods, ML adopts a data-driven approach that directly learns the underlying patterns and relationships from data without explicitly deriving the mathematical formulations of the physical processes involved.

In the application of predicting grain growth evolution, ML offers a number of advantages. Neural networks, a subset of ML, excels in detecting latent patterns and learning complex evolutionary patterns in the grain growth mechanism. Such networks are capable of capturing both explicit and implicit relationships that drive the mechanism of grain growth. This is particularly important when considering the computational efficiency as it could predict evolutions at speeds which are potentially orders of magnitude faster than the conventional simulation methods, primarily due to bypassing the numerous steps in the conventional simulation such as solving differential equations and iterative updates. This improvement leads to applications in real-time processing and large-scale simulations. Moreover, ML approaches demonstrate notable scalability in comparison with the conventional simulation methods. In the conventional simulation methods, the computational burden tends to scale exponentially with system size, yet the ML approaches allow for larger system sizes without an equally exponential computational cost. Furthermore, the ML approaches exhibit considerable adaptability as they can learn from both experimental and simulated data, and could potentially capture other subtle phenomena that theoretical models might overlook.

Recent advancements in the application of ML to accelerate the prediction of microstructure evolution have shown interesting results. The authors in \cite{sase_prediction_2023} proposed a neural networks to predict the evolution of atomic-scale microstructures, which combines a dimensionality reduction model—Variational Autoencoder (VAE)—with Long Short-Term Memory (LSTM), a neural networks architecture designed to learn the temporal patterns from previous timestep. The architecture consists of a three-stage pipeline including dimensionality reduction using encoder block of VAE into ten-dimensional (10-D) latent variables, predicting the temporal evolution with LSTM, and then reconstructing the predicted evolution back to the original dimensions using decoder of the VAE. As a result, they were able to accurately predict the evolution of polycrystalline iron microstructures in a fraction of the time required by the conventional simulation.

In \cite{ahmad_accelerating_2023}, the combination of Autoencoders with Convolutional LSTM (ConvLSTM) is explored, leveraging both spatial and temporal feature extraction capabilities to accelerate the prediction of spinodal decomposition. After training on microstructure data generated from phase-field simulations of known compositions, the model successfully predicted future microstructural states based on previous timestep for new compositions. Their proposed method resulted in a significant reduction in computational time in predicting microstructure evolution while maintaining high-resolution representations.

Similarly, in \cite{Sun2023}, a computationally efficient surrogate model that directly learns microstructural evolution by incorporating phase-field simulations with ML methods is developed. Their approach combined low-dimensional representation of microstructure with time-series models LSTM. The trained neural networks acted as a surrogate model in which intermediate outputs from the phase-field simulation at a specific timestep are used to predict the remaining evolution. Their model demonstrated superior performance in accurately predicting the nonlinear dynamics of spinodal decomposition within seconds. This outcome significantly addressed the computational cost associated with full-scale phase-field simulations.

These advancements above highlight the capability of ML, specifically neural networks, to successfully learn and generalize complex microstructural evolution patterns while substantially reducing computational costs in comparison to the conventional simulation methods. However, while these ML methods show promising results in mechanisms like spinodal decomposition and atomic-scale evolution, their application in predicting grain growth remains unexplored. Grain growth set distinct challenges in comparison to other mechanisms due to its multi-scale nature, complex boundary interactions, and the need to maintain topological consistency over extended temporal scales. In addition, simulating grain growth evolution with predictive high-fidelity formulations takes days or weeks to complete, placing a significant bottleneck in materials design and optimization pipelines, leading to impracticality in industrial context.

These challenges are the motivation behind this research. This research aims to focus on developing an ML-based approach that accelerates the prediction of grain growth while maintaining acceptable levels of accuracy on par with conventional simulation methods. Three interconnected aspects drives the direction of this study: first, to explore and validate the applicability of ML in the task of predicting (in a quantitative manner) the grain growth evolution; second, to address the computational challenges inherent in the conventional simulation methods; and third, to meet the growing industrial demands for rapid, and accurate predictions of grain growth behavior while balancing the computational efficiency.

This research, thus, represents a step toward integrating advanced computational methodologies with materials science principles. Developing ML techniques for accelerating grain growth prediction would also serve as a template for addressing similar challenges in other mechanisms, such as recrystallization or Smith-Zener pinning \cite{BERNACKI2024101224}, potentially catalyzing a broader adoption of ML approaches in materials design and optimization processes.

\section{Methodology}
\label{Methodology}
In this study, the grain growth evolution is modeled as a spatio-temporal learning task in ML. The spatial aspect refers to how grains and grain boundary networks are arranged within the domain, whereas the temporal aspect describes how these arrangements evolve over time. To address this complex learning task, the proposed methodology consists of three interconnected components: first, generating a comprehensive dataset of the grain growth evolution using validated simulation tools; second, building a hybrid neural network specifically designed for predicting the grain growth evolution; and third, implementing rigorous evaluation metrics to assess the model's performance.

\subsection{Dataset}
The dataset required for this study are sequences of microstructure images that capture evolution of the grains and the grain boundaries at different temporal points. To generate comprehensive and diverse evolution sequences for model training, we utilized validated in-house tools. The dataset generation pipeline followed a two-step process: first, LavoGen \cite{florez2020novel} generated the initial microstructure states using a specific set of input parameters; then, this initial state was fed into ToRealMotion (TRM) \cite{florez2020novel} to simulate thermal treatment with another set of chosen parameters. TRM has been developed and validated for grain growth modeling, ensuring accuracy and reliability that reflects real-world evolution. Table \ref{tab:dataset_gen_parameters} describes the input parameters which were chosen to accurately represent the physical mechanism of interest and ensure that the generated evolutions reflect real materials behavior.

The initial microstructure state generated by LavoGen in the form of finite-element (FE) mesh based on the specific parameters are listed in Table \ref{tab:dataset_gen_parameters}. These parameters including the size of domain ($a$), type of grain size distribution, mean grain size ($\overline{R}$), and standard deviation ($\sigma$), each plays a role in shaping the initial microstructure. The size of domain determines the overall spatial extent of the simulated microstructure, influencing total number of grain and their interactions. The type of grain size distribution affects how grain sizes are statistically spread around the mean, impacting the heterogeneity of the microstructure, with Normal distribution ($\mathcal{N}$) being the primary focus for this study. The mean grain size sets the average size of grains in a domain which influences overall material properties, while standard deviation controls the variability in grain sizes around the mean, with higher values resulting in greater spread in grain sizes.

Once the initial microstructure was generated, it was then passed to TRM with another set of parameters (Table \ref{tab:dataset_gen_parameters}). TRM is a simulation tool that is based on front-tracking method to simulate microstructure evolution during hot metal forming at polycrystalline scale. The parameters for the simulations include annealing temperature ($T$), reduced mobility ($\mu\gamma$), and annealing duration. The treatment temperature sets which thermal conditions for the simulation to run, directly influencing grain growth kinetics. Higher temperatures typically accelerate grain boundary movement, while lower temperatures result in slow grain boundary movement. Reduced mobility refers to the inherent resistance/drag to grain boundary movement, affecting the overall growth rates. It is important to note that the chosen value for the reduced mobility parameter resulted in dimensionless mobility, compared to real-world evolution where mobility values would be tied to specific material properties and processing conditions. This abstraction to dimensionless mobility represents a key advantage of our approach as it allows us to process the microstructure evolution data in a normalized framework that can be more easily transferred across different material systems in downstream tasks. Last but not least, treatment duration specifies the total simulation time, impacting the extent of observable microstructure evolution, with longer durations generally leading to more pronounced grain growth and coarsening. The used data are representative for annealing applied to a large variety of metallic alloys. Most importantly, it is worth highlighting the versatility of the model once trained on this data. Indeed, a large initial difference in average grain size can be accounted for by normalizing the grain size data by the average grain size. Similarly, differences in temperature and/or reduced mobility can be taken into account by stretching the model's time scale after training. In other words, for example, the microstructural state reached at a given time $t$ by a polycrystal with twice the mobility will be the same as the initial prediction for that polycrystal at time $2t$. The choice of the initial form of the grain size distribution should not be a concern regarding the sufficient representativeness of the distributions seen during model training. Indeed, it is well known that a transition phase toward log-normal-type distributions automatically occurs during isotropic grain growth. This transition inherently ensures a certain diversity in the evolving microstructures used for model training.

\begin{figure}[htbp]
    \centering
    \includegraphics[width=0.8\textwidth]{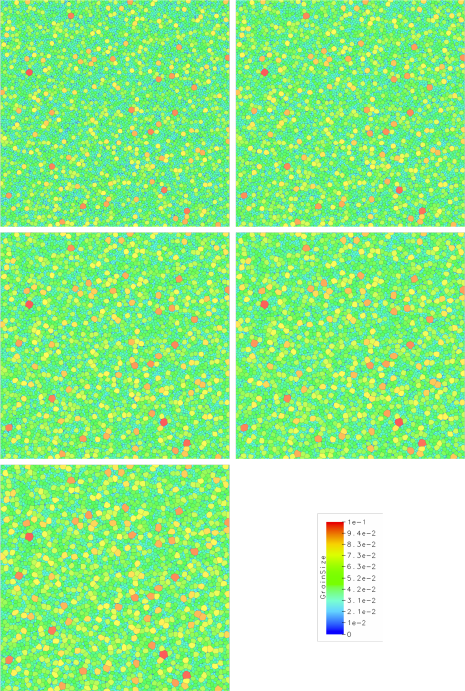}
    \caption{Temporal evolution of a $\SI{5}{\milli\meter}\times\SI{5}{\milli\meter}$ microstructure. From top to bottom and left to right: $t=\SI{0}{\minute}$ (initial state), $t=\SI{15}{\minute}$, $t=\SI{30}{\minute}$, $t=\SI{45}{\minute}$, and $t=\SI{1}{\hour}$. The color code corresponds to the equivalent circle radius (ECR) in mm. The ECR of a grain is estimated as the radius of the circle having the same area as the considered grain.}
    \label{fig:microstructure_evolution}
\end{figure}

To increase the size of the generated dataset and enhance the diversity of the evolutions, several data augmentation techniques were applied to improve generalization and capture invariant features for model training. These techniques include rotation of each microstructure image at multiple angles, varying scaling factors to simulate different magnification levels, and random image cropping to focus on different regions of microstructure. As a result, the final dataset consisted of 648 sequences, each showcasing isotropic grain growth over 1 hour. Figure \ref{fig:microstructure_evolution} illustrates an example of the generated evolutions at various timesteps, with the configuration $\overline{R}_{t=0s}=\SI{20}{\micro\meter}$, $D=\SI{5}{\milli\meter}$, $\sigma_{t=0s}=\SI{16}{\micro\meter}$. The figure clearly demonstrated grain growth patterns, boundary movements, and structural transformations, serving as key data for the neural networks to learn patterns of the underlying mechanism. In terms of computational expense, the simulation of TRM based on the conventional methods took approximately 6 hours to produce 18 raw sequences on a single computing machine with a recent 16-core CPU, with each sequence requiring about 20 minutes on average to complete. It is important to note that while TRM supports distributed computation \cite{Florez2020c}, the simulations conducted in this study were performed sequentially on a single machine without parallelization.

\begin{table}[htbp]
\centering
\footnotesize
\renewcommand{\arraystretch}{1.2} 
\setlength{\extrarowheight}{2pt} 
\begin{tabularx}{\textwidth}{p{2.4cm}p{2.2cm}>{\centering\arraybackslash}p{2.5cm}X}
\toprule
\textbf{Tool} & \textbf{Parameter} & \textbf{Value(s)} & \textbf{Description} \\
\midrule
\multirow{4}{=}{Initialization} & $a$ (mm) & $2$, $3$, $4$, $5$ & Side length of the square domain in millimeters. \\ \cline{2-4}
 & Distribution & $\mathcal{N}(\overline{R},\,\sigma^{2})$ & The type of distribution used for the ECR distribution (Laguerre-Voronoï tesselation \cite{BERNACKI2024101224,Hitti2012}). \\ \cline{2-4}
 & $\overline{R}$ $(\SI{}{\micro\meter})$ & $20$ & The mean value of the ECR distribution. \\ \cline{2-4}
 & $\sigma$ $(\SI{}{\micro\meter})$ & $2$, $4$, $8$, $16$, $32$ & The standard deviation of the ECR distribution. \\ 
\midrule
\multirow{3}{=}{Annealing simulation} & $T$ $(\SI{}{\kelvin})$ & $1323.15$ & The absolute temperature. \\ \cline{2-4}
 & $\mu\gamma$  $(\SI{}{\square\milli\meter\per\second})$ & $1 \times 10^{-6}$ & The reduced mobility. \\ \cline{2-4}
 & Annealing duration ($\SI{}{\minute}$) & $60$ & The physical time of the simulation.  \\
\bottomrule
\end{tabularx}
\caption{Parameters for constructing a dataset using TRM simulation tools.}
\label{tab:dataset_gen_parameters}
\end{table}

\subsection{Model Architecture}
\begin{figure}[htbp]
    \centering
    \includegraphics[width=1\linewidth]{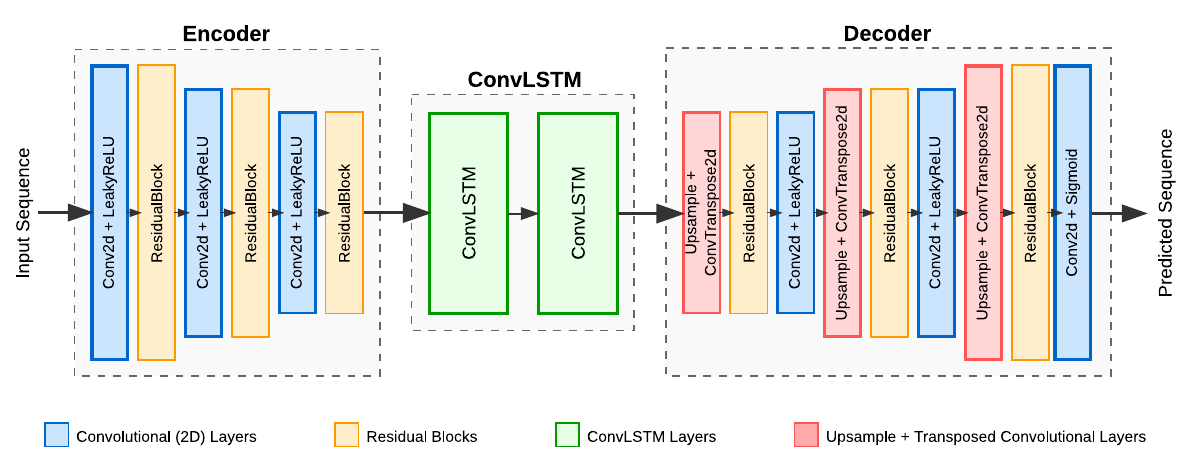}
    \caption{Neural network architecture for predicting subsequence grain evolution, consisting of an encoder-decoder autoencoder with residual blocks integrated with ConvLSTM layers.}
    \label{fig:nn_architecture}
\end{figure}

Building on the recent advancements \cite{sase_prediction_2023,ahmad_accelerating_2023,Sun2023}, a hybrid neural networks architecture was adopted by combining an Autoencoder, to handle spatial dimensionality reduction and reconstruction, with ConvLSTM, for spatio-temporal pattern learning, as illustrated in Figure \ref{fig:nn_architecture}. 

Autoencoder is a type of neural networks \cite{hinton-2006}, generally consisted of encoder and decoder blocks. Encoder block serves as a dimensionality reduction model which encodes input data into a compact representation in the latent space. On the other hand, decoder block reconstructs the compressed data to its original dimensions for the prediction output. 

In this study, the encoder block consisted of Convolutional layers to reduce the dimension of the data and remove irrelevant features that do not contribute to the model learning, creating a bottleneck learning effect where only the most relevant features of the mechanism presented in the sequences were passed through to ConvLSTM for pattern learning and prediction. Conversely, the decoder block reconstructed the predicted evolution from ConvLSTM through Upsampling layers, Transposed Convolutional layers, and Convolutional layers. The Upsampling and the Transposed Convolutional layers progressively increased spatial dimensions from the compressed dimension, while the Convolutional layers refined the reconstructed features by enhancing local details, resulting in high-fidelity microstructure representation. On top of that, the Autoencoder incorporates Residual Blocks which could potentially address the problem of gradient vanishing and enhance feature preservation during dimensionality reduction and reconstruction processes. 

In between the encoder block and decoder block, ConvLSTM layers handle the spatio-temporal pattern learning for the mechanism. The ConvLSTM is a variant of the Recurrent Neural Networks \cite{shi2015convolutional}, capable of learning long-term dependencies by incorporating convolution operators at each gates in the LSTM cell, making it well-suited for processing spatio-temporal data like grain growth evolution. Unlike LSTM which processes vector sequences, ConvLSTM operates on five-dimensional (5-D) tensors, instead of three-dimensional (3-D) tensor found in LSTM, in order to preserve spatial correlations while model temporal patterns. In this study, layers of the ConvLSTM processed the compressed features from the encoder block of the Autoencoder to capture both local and global spatio-temporal dependencies in the grain growth evolution mechanism. Specifically, these layers learned how grain boundaries move and interact over time, detecting patterns of grain evolution and predicting the growth between neighboring grains. The networks learned to recognize when certain grains will grow at the expense of others based on their size, shape, and relative positions, allowing for accurate multi-step predictions of the evolution. Throughout these layers, the spatial structure and temporal patterns of the input were maintained while continuously predicting future state of the evolutions in a compressed form, which were then passed to the decoder block of the Autoencoder for reconstructing the result.

This hybrid architecture fits with the spatio-temporal task of predicting grain growth while maintaining the physical validity of grain boundaries and topology.

\subsection{Evaluation Framework}
\subsubsection{Loss Function}
To ensure optimal learning of the spatial and temporal features of grain growth evolution, a customized adaptive weight loss function was developed as:

\begin{equation}
    L_{total} = \alpha(e)L_{MSE} + \beta(e)L_{SSIM} + \gamma(e)L_{BP},
\end{equation}

where $L_{MSE}$ is the Mean Squared Error (MSE) \cite{10.1214/009053604000000201} loss, $L_{SSIM}$ is the Structural Similarity Index Measure (SSIM) \cite{hore-2010} loss, $L_{BP}$ is the Boundary Preservation loss (BP) \cite{kervadec2019boundary}, and $\alpha(e)$, $\beta(e)$, and $\gamma(e)$ are the epoch-dependent weighting coefficients where $\alpha$ + $\beta$ + $\gamma$ = 1. This adaptive weighting scheme optimized the learning process across different stages of training. The weighting coefficients evolved throughout the training process following a scheduled progression:

\begin{align*}
    \alpha(e) &= \max(1 - \max(e-T, 0)/(E-T), 0.5) \\
    \beta(e) &= \min(\max(e-T, 0)/(E-T), 0.25) \\
    \gamma(e) &= \min(\max(e-T, 0)/(E-T), 0.25)
\end{align*}

where $e$ is the current epoch, $T$ is the initial epochs at which the transition begins, and $E$ is the epochs by which the transition should be completed. During the initial training phase, the model was prioritized to learn the evolution patterns and reconstruct overall grain boundaries and topology by setting $\alpha(0) = 1$, $\beta(0) = 0$, and $\gamma(0) = 0$. As training progressed, the weight of $\beta(e)$ and $\gamma(e)$ were gradually increased, while $\alpha(e)$ were decreased, but never drops below $0.5$, in order to enhance reconstruction quality in terms of perceptual fidelity and structural similarity. 

The $L_{MSE}$ is a pixel-wise loss function which computed the mean squared pixel-wise difference between the predicted and ground truth microstructure. This loss component directly influenced the accuracy of the model in terms of numerical representation in predicting pattern and reconstructing grain morphologies.

In addition, the $L_{SSIM}$ served as an additional loss component for evaluating human perceptual features such as local patterns, texture, luminance, contrast, and structural similarity between the predicted and ground truth microstructure images. This loss was particularly beneficial in the context of grain microstructures as it helped preserves the critical spatial relationships and textural characteristics, such as structure of grain boundary networks and structure of multiple junctions, that are essential in predicting the evolution from timestep to timestep.

As stated, grain boundary is a crucial feature in the task of predicting grain growth evolution. In addition to the $L_{MSE}$ and the $L_{SSIM}$, the $L_{BP}$ was another important loss component for further preserving grain boundary features by capturing fine structural details, such as grain boundaries and multiple junctions, that go beyond the human perceptual similarities captured by $L_{SSIM}$ in predicted images. This helped the model accurately reconstruct the complex microstructural patterns essential to grain growth. 

The proposed multi-component loss function with the adaptive weighting scheme provides a comprehensive framework that balance pixel-wise accuracy, overall structural similarity, and finer details of the grain boundaries and junctions for training the neural network. Consequently, the outcome achieved a level of accuracy where the prediction from the trained model could potentially serve as reliable inputs for downstream tasks such as post-prediction analysis or further processing.

\subsubsection{Training Hyperparameters}
\begin{table}[htbp]
    \centering
    \begin{tabularx}{0.66\textwidth}{l|c}
    \toprule
    \textbf{Parameter} & \textbf{Value} \\
    \midrule
    Learning rate & $1 \times 10^{-4}$ \\
    Optimizer & AdamW \\
    Temporal window size ($\SI{}{\minute}$) (Input-Predict) & ${10-10}$; ${20-20}$; ${30-30}$ \\
    Dataset split ratio ($\SI{}{\percent}$)(Train/Val/Test) & $70/20/10$ \\
    Batch size & $1$ \\
    Epochs & $60$ \\
    Gradient clipping threshold & $1.0$ \\
    \bottomrule
    \end{tabularx}
    \caption{Hyperparameters for model training.}
    \label{tab:training_hyperparams}
\end{table}

In this task, hyperparameters also played a crucial role in model training. The selection of the parameters was guided by a combination of grid search experimentation and empirical testing to ensure optimal performance for our specific dataset and the adopted architecture. Table \ref{tab:training_hyperparams} listed the key hyperparameters used in the final training configuration.

The learning rate of $1\times 10^{-4}$ was chosen after experimenting with several different learning rates ranging from $1\times 10^{-6}$ to $1\times 10^{-3}$. The chosen learning rate provided an optimal balance between convergence speed and training stability. Higher learning rates resulted in training instability, particularly in capturing the temporal evolution of grain boundary networks, while smaller learning rates caused excessively slow convergence.

AdamW optimizer \cite{loshchilov2017decoupled}, a variant of Adam optimizer\cite{kingma2014adam} that decouples weight decay from the gradient update, was chosen as optimizer for the training. AdamW has shown superior performance in deep learning training in tasks by preventing the optimizer from being overly influenced by noisy gradients, which is particularly important when modeling complex microstructural evolution processes with inherent stochasticity.

Concerning temporal windows, a number of temporal window sizes were tested to determine which provided the optimal prediction accuracy. We established a naming convention of $S-XX-YY$ for each model, where $XX$ represents the input sequence length and $YY$ represents the target prediction sequence length (both in minutes). For instance, S-10-10 means the model takes 10 minutes of evolution of as input data to predict the subsequent 10 minutes of evolution. In this study, the temporal windows tested were S-10-10, S-20-20, and S-30-30, representing different temporal window sizes. This test was insightful as sequence-to-sequence models can be sensitive to the temporal window size, with shorter sequences potentially missing long-term dependencies and longer sequences potentially introducing unnecessary complexity and computational overhead.

Regarding the data distribution, the dataset was split in a \SI{70}{\percent}, \SI{20}{\percent}, \SI{10}{\percent} ratio for training, validation, and testing, respectively. This split ratio was carefully designed to ensure that each sequence in the generated dataset, along with its augmented versions, was presented across all split sets to maintain diverse sequences for balanced training and evaluation. Additionally, to ensure balanced temporal representation, the split method involved distributing different temporal windows evenly across all sets. For example, each set contained early temporal windows, middle temporal windows, and last temporal windows of the sequences. This approach prevented potential bias from grouping similar windows together and guaranteed that each split contained a diverse mix of window types.

Gradient clipping with a threshold of $1.0$ proved to be crucial due to the multi-scale nature of grain evolution where sudden boundary movement changes could cause gradient explosions. This threshold was determined by monitoring gradient norms during initial training runs, setting it just above typical gradient magnitudes to prevent extreme updates while allowing normal training progression.

Each model, with different temporal window size configuration, was trained for 60 epochs on an in-house computing platform equipped with an NVIDIA A100 GPU. Depending on size of the temporal window, the training process took approximately 2 to 4 hours per configuration to complete. This lengthy process was necessary to ensure that the model adequately learned the complex relationships between initial microstructural states and their subsequent evolution while avoiding overfitting.

\subsubsection{Evaluation Metrics}
To evaluate the performance of the trained models, various evaluation metrics were employed including adapted pixel-wise accuracy to evaluate the grain boundary networks, as well as perceptual and structural similarity measures. Moreover, to measure the accuracy in terms of physical characteristics, statistical metrics such as the error in mean grain size, Kullback–Leibler (KL) divergence \cite{kullback1951information}, and Wasserstein distance \cite{villani2008optimal} were also calculated. 

Adapted versions of MSE and Mean Absolute Error (MAE), called Boundary-Focus MSE ($MSE_{b}$) and Boundary-Focus MAE ($MAE_{b}$), were computed as in Equations \ref{eq:adapated_mse} and \ref{eq:adapated_mae}, respectively. Unlike standard MAE and MSE \cite{hyndman-2006, 10.1214/009053604000000201} which calculate the error across all pixels in the microstructure images, the Boundary-Focus MSE/MAE calculated only the grain boundary network between predicted and ground truth in an image. In the context of grain growth, this boundary-focused metric provides a more relevant evaluation as reconstructed predicted images could potentially contain artifacts outside grain boundary network. Hence, this adapted version of MSE and MAE resulted in an accurate evaluation of the model's ability to predict grain boundary network while discarding the artifacts.

\begin{equation}
    MSE_{b} = \frac{1}{N} \sum_{i=1}^{N} (B_{pred} - B_{true})^2,
    \label{eq:adapated_mse}
\end{equation}

\begin{equation}
    MAE_{b} = \frac{1}{N} \sum_{i=1}^{N} |B_{pred} - B_{true}|.
    \label{eq:adapated_mae}
\end{equation}

In these equations, $B_{pred}$ and $B_{true}$ denote grain boundary networks of predicted and ground truth which was extracted using Sobel edge detection algorithm \cite{vincent2009descriptive}, while $N$ is the total number of pixels in images.

To assess perceptual quality and structural similarity, Peak Signal-to-Noise Ratio (PSNR) and SSIM \cite{hore-2010} were employed. PSNR provides a quantitative measure of reconstruction quality. Higher PSNR values indicated better image fidelity, lower levels of noise, and distortion in the predicted images. This metric is useful for evaluating how well the fine details and edges, particularly at grain boundaries, are produced compared to the ground truth.

In contrast, SSIM evaluates image similarity based on structural information and correlates more closely with human visual perception as discussed previously. An SSIM value close to 1 indicates high structural similarity, suggesting that the predicted image closely matches the ground truth, while values closer to 0 reflect poor structural similarity. When applied to predict grain images, SSIM effectively captures whether the overall topology of the grain boundary network and the grain size distribution are preserved. This ensures that the predictions not only appear visually similar to the ground truth but also retain the functionally important structural characteristics of the microstructure.

In terms of statistical metrics, the grain size distribution was computed by applying image processing techniques, such as segmentation, to the predicted and ground truth images. From these distributions, several error measures were calculated, including the mean grain size error ($\overline{R}$), the KL divergence, and the Wasserstein distance. 

The mean grain size error quantifies the relative difference in average grain size between the predicted and ground truth images, providing insight into how accurately the model preserves overall grain growth trends. It also reflects whether the shape of the grain size distribution is correctly captured in the predictions.

KL divergence quantifies how much the predicted probability distribution diverges from the ground truth distribution. It provides insight into the information loss when using the predicted distribution to approximate the actual one. A lower KL divergence indicates that the predicted grain size distribution closely matches the ground truth in terms of overall shape and probability allocation.

In addition to KL divergence, the Wasserstein distance, also known as the Earth Mover’s Distance, was used. This metric measures the minimum cost of transforming one distribution into another, where the cost is defined as the amount of probability mass moved multiplied by the distance it is moved. Unlike KL divergence, Wasserstein distance offers the advantages of being symmetric and more robust when distributions have different supports or minimal overlap, making it particularly valuable for comparing multi-modal grain size distributions. This metric is particularly valuable in grain size analysis because it accounts for both the shape differences and the magnitude of deviation in grain sizes. As a result, it provides a physically interpretable measure of how far the predicted distribution deviates from the ground truth in the actual grain size.

This sophisticated approach is crucial as high visual or pixel-wise accuracy does not inherently guarantee an accurate representation of the material’s physical properties. In some cases, a model may achieve a very high pixel-level accuracy while failing to correctly capture the underlying grain size distribution mainly for small-sized grains. This can lead to significant errors when such predictions are used to estimate material properties or in downstream microstructure–property relationship studies. Therefore, evaluating both perceptual quality and statistical accuracy ensures that the predictions not only appear visually similar to the ground truth but also maintain physical fidelity in terms of grain morphology and distribution, features that directly influence the mechanical properties of the material.

\section{Results}
\label{Result}
Our aim in developing this accelerated method is to predict the final state of the isotropic grain growth evolution by using early evolutions from the conventional TRM simulation. Specifically, each model was provided with the required number of input frames, and then made a prediction. The prediction was then used as input for the next iteration, and continuing until the final prediction point is reached depending on temporal window size of the trained model. The predicted results were then benchmarked, with metrics discussed previously, against the conventional TRM simulation, which served as our ground truth.

\begin{figure}[htbp]
    \centering
    \begin{subfigure}[b]{0.48\textwidth}
        \centering
        \includegraphics[width=\textwidth]{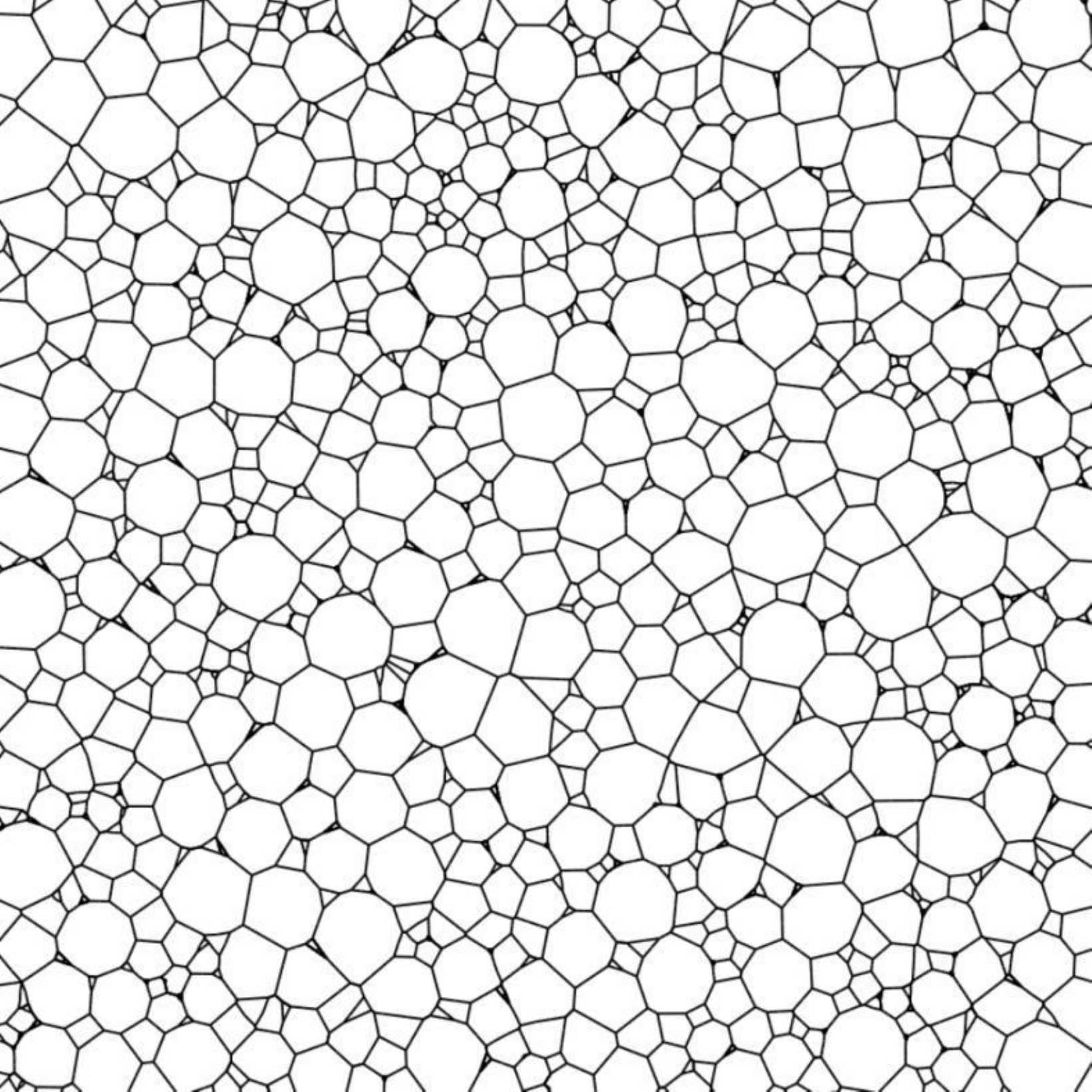}
        \label{fig:subfig1}
    \end{subfigure}
    \hfill
    \begin{subfigure}[b]{0.48\textwidth}
        \centering
        \includegraphics[width=\textwidth]{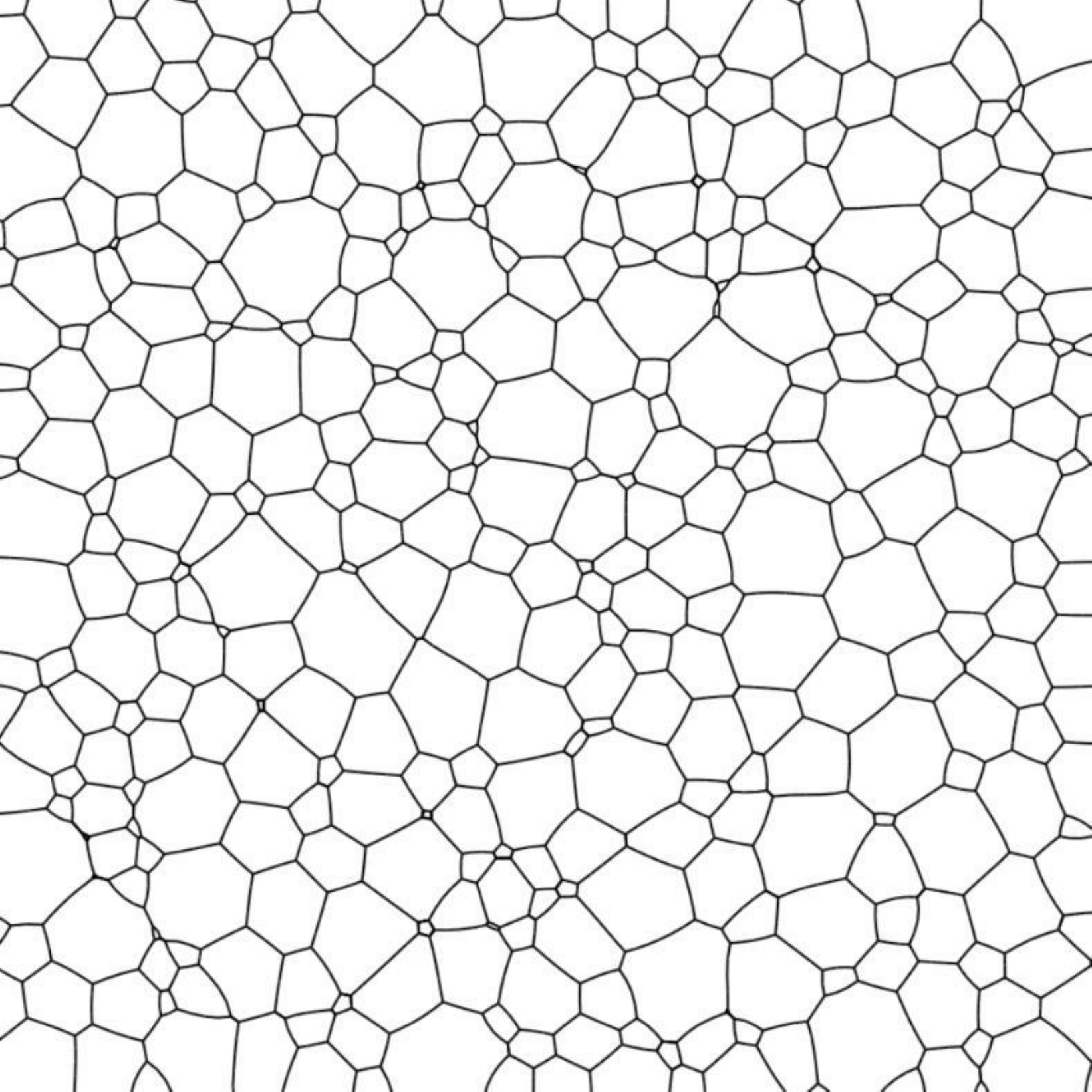}
        \label{fig:subfig2}
    \end{subfigure}
    \caption{Visualization of a prediction case (a) Microstructure state at $t=\SI{0}{\hour}$ generated by LavoGen, serving as input to the conventional TRM simulation and the ML models. (b) Microstructure state at $t=\SI{1}{\hour}$ predicted by the conventional TRM simulation, containing approximately 300 grains evolved from an initial 800 grains, serving as the ground truth reference for evaluating ML model predictions.}
    \label{fig:input_prediction}
\end{figure}

The trained models were extensively tested across a range of complexity levels, from simpler configurations to more extreme cases. The results and findings remained consistent across this range of test. For this study, we chose a prediction case on a $\SI{2}{\milli\meter}\times \SI{2}{\milli\meter}$ domain and predicted 1-hour evolution of grain growth as an optimal and representative scenario, as illustrated in Figure \ref{fig:input_prediction}. Statistically, the chosen case contained approximately 800 grains with complex grain junctions that evolve and reduced to slightly above 300 grains after 1 hour of evolution via the conventional TRM simulation. This final state and its statistical information served as the reference for assessing our trained models' performance. Given this scenario, the S-10-10, which processed 10-minute input sequences, required 5 iterations to reach the final time, whereas the S-30-30 required only 1 iteration. 

\begin{table}[htbp]
    \centering
    \footnotesize
    \setlength{\tabcolsep}{3pt}
    \begin{tabularx}{0.675\textwidth}{l|cccc|ccc}
    \toprule
    \multirow{2}{*}{\textbf{Model}} & \multicolumn{4}{c|}{\textbf{Pixel-wise and Perceptual Accuracy}} & \multicolumn{3}{c}{\textbf{Statistical Accuracy}} \\
    \cmidrule(lr){2-5} \cmidrule(lr){6-8}
    & \textbf{$\textbf{MSE}_\textbf{b}$} $\downarrow$ & \textbf{$\textbf{MAE}_\textbf{b}$} $\downarrow$ & \textbf{PSNR} $\uparrow$ & \textbf{SSIM} $\uparrow$ & \textbf{$\overline{R}$ Error (\%)} $\downarrow$ & \textbf{KL Div.} $\downarrow$ & \textbf{W} $\downarrow$ \\
    \midrule
    S-10-10 &  0.1509 & 0.3265 & 14.2984 & 0.6950 & 2.2708 & 0.0756 & 0.0021 \\
    S-20-20 &  0.0999 & 0.2607 & 18.4541 & 0.8289 & 0.4174 & 0.0069 & 0.0004 \\
    S-30-30 &  0.0680 & 0.2067 & 19.9589 & 0.8671 & 0.0708 & 0.0048 & 0.0003 \\
    \bottomrule
    \end{tabularx}
    \begin{tablenotes}
    \footnotesize
    \centering
    \item KL Div. = KL divergence (predicted to ground truth); W = Wasserstein distance
    \end{tablenotes}
    \caption{Performance comparison of trained models across different temporal window size.}
    \label{tab:model-performance}
\end{table}

Table \ref{tab:model-performance} summarizes the performance of different trained models across various evaluation metrics as previously discussed. Notably among all, the S-30-30 outperformed the other models across metrics from pixel-wise/perceptual to statistical accuracy. In contrast, the S-10-10 consistently recorded the lowest performance across metrics, while the S-20-20 achieved results closer to the S-30-30. Despite being the low performing model, the S-10-10 still maintained a relatively low mean grain size error of about \SI{2.3}{\percent}, while the other two models achieved errors below \SI{0.5}{\percent}. Furthermore, both the S-20-20 and the S-30-30 predicted grain size distributions that closely matched the ground truth, as proven by the low KL divergence and Wasserstein distance values.

\begin{figure}[htbp]
  \begin{subfigure}{0.32\textwidth}
    \includegraphics[width=\textwidth]{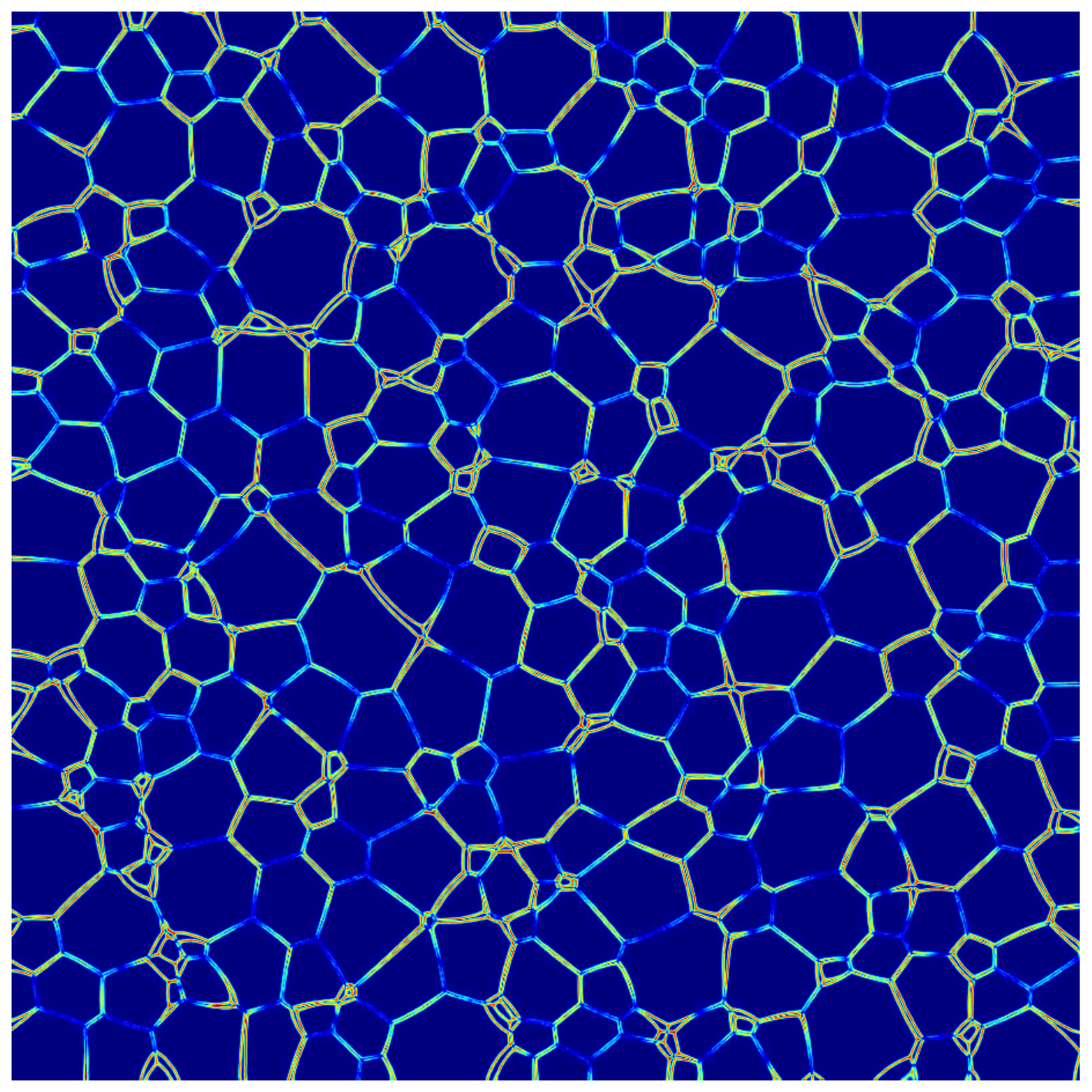}
    \caption{}
  \end{subfigure}
  \hfill
  \begin{subfigure}{0.32\textwidth}
    \includegraphics[width=\textwidth]{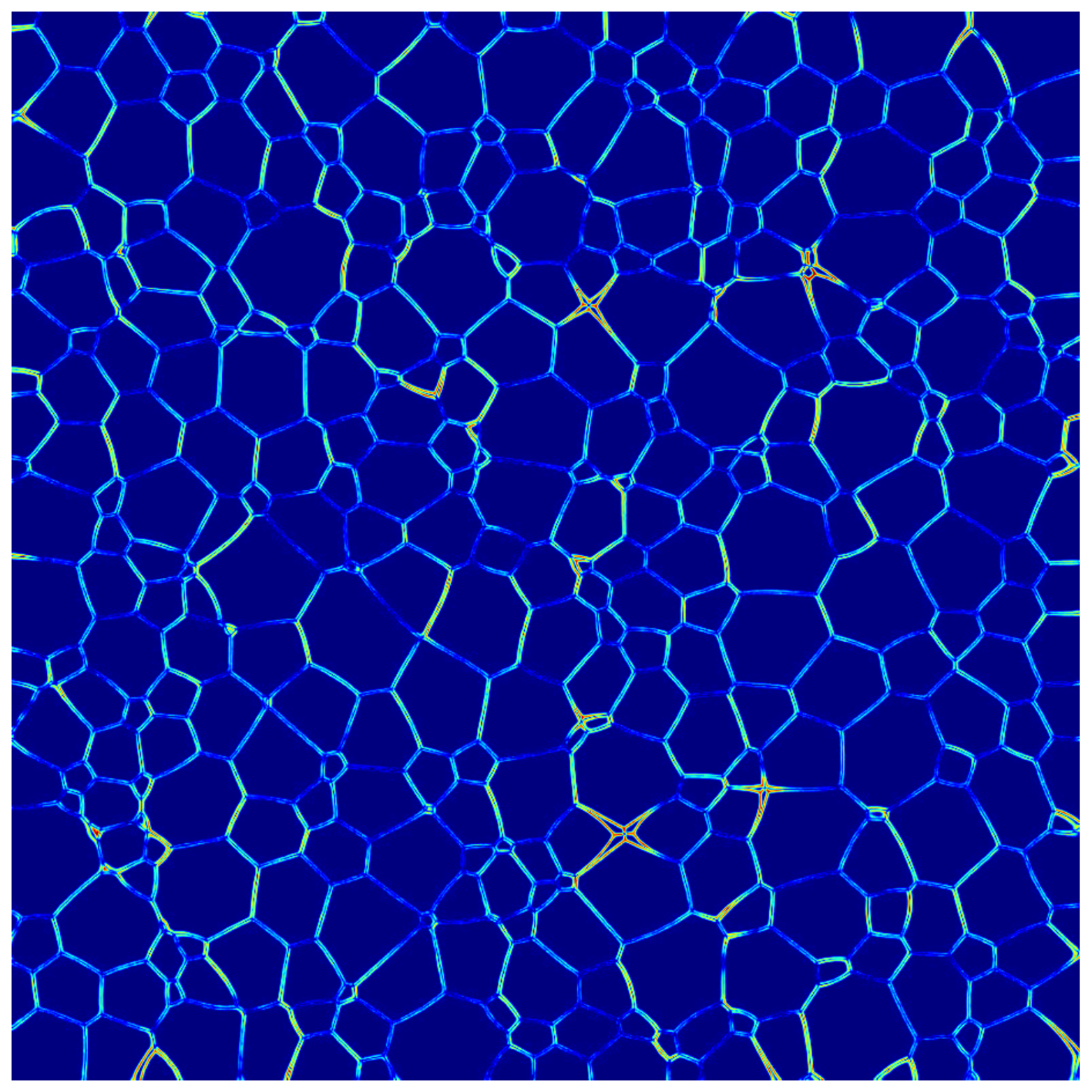}
    \caption{}
  \end{subfigure}
  \hfill
  \begin{subfigure}{0.32\textwidth}
    \includegraphics[width=\textwidth]{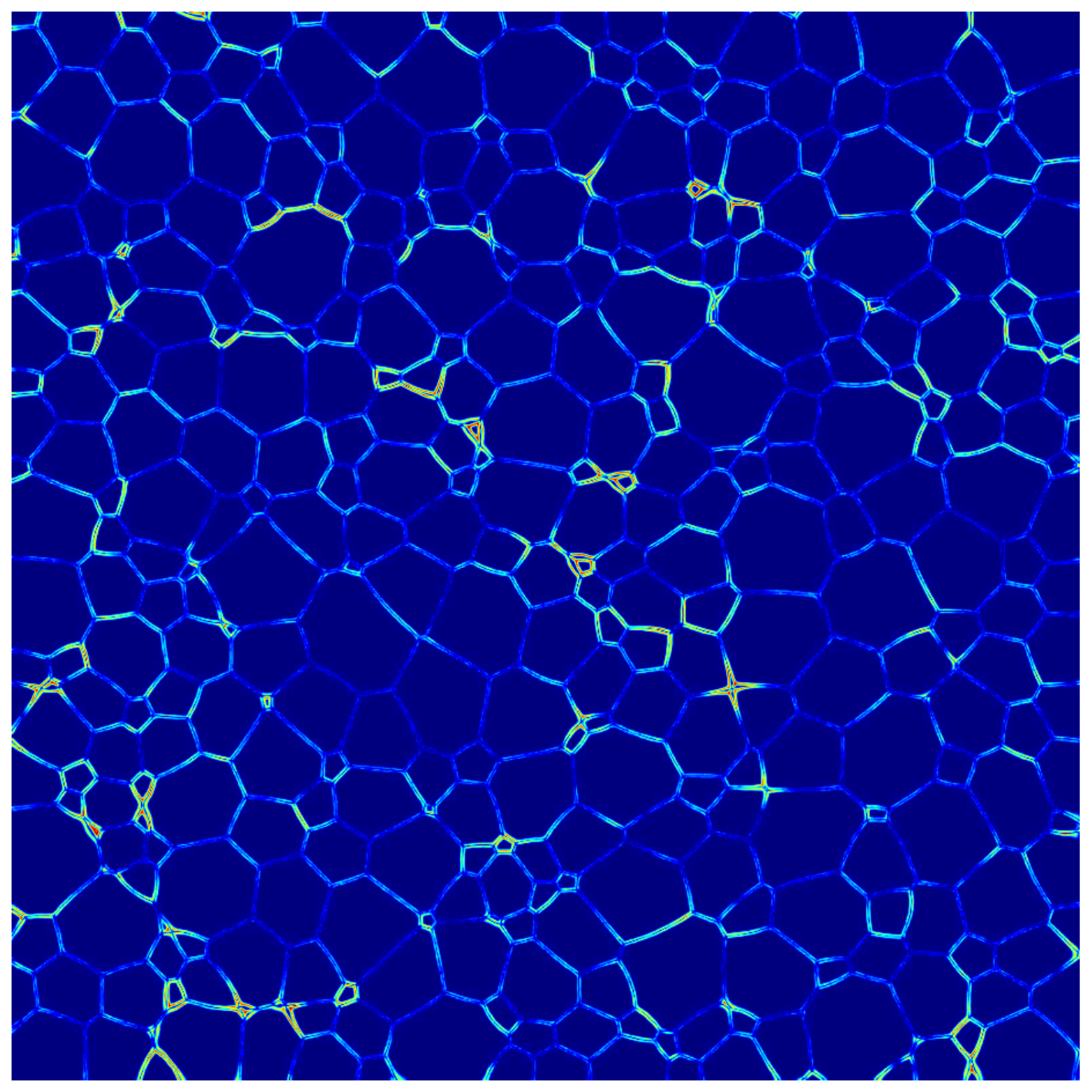}
    \caption{}
  \end{subfigure}
  \vspace{0.10cm}
  \centering
  \includegraphics[width=1\textwidth]{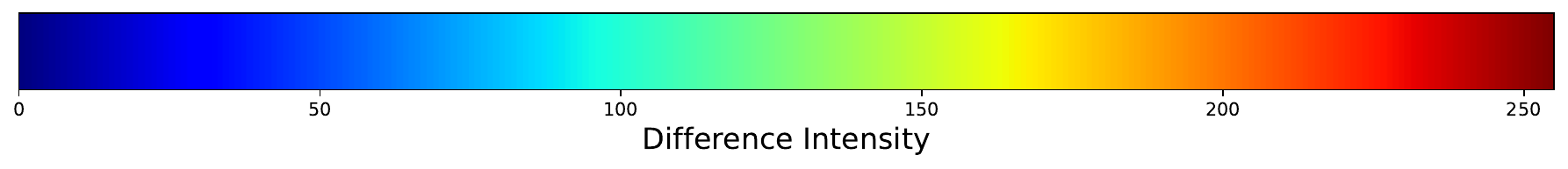}
  
  \caption{Heatmap visualization of prediction accuracy across different model configurations. (a) S-10-10 model. (b) S-20-20 model. (c) S-30-30 model. The common color scale indicates the magnitude of prediction error.}
  \label{fig:heatmaps}
\end{figure}

In terms of structural similarity, both the S-20-20 and the S-30-30 achieved over \SI{80}{\percent} of SSIM scores, indicating a close match to the grain boundary networks produced by the conventional TRM simulation. Figure \ref{fig:heatmaps} visualizes the deviations in grain boundary topology at $t=\SI{1}{\hour}$. Red regions indicate higher deviations, while blue regions denote strong alignment with the ground truth. From the figure, it is evident that the S-10-10 model introduces the largest errors, especially at grain junctions. In contrast, the S-30-30 exhibits the smallest differences. These visual trends align with the quantitative metrics reported in Table \ref{tab:model-performance}.

\begin{figure}[htbp]
    \centering
    \begin{subfigure}[b]{0.49\textwidth}
        \centering
        \includegraphics[width=\textwidth]{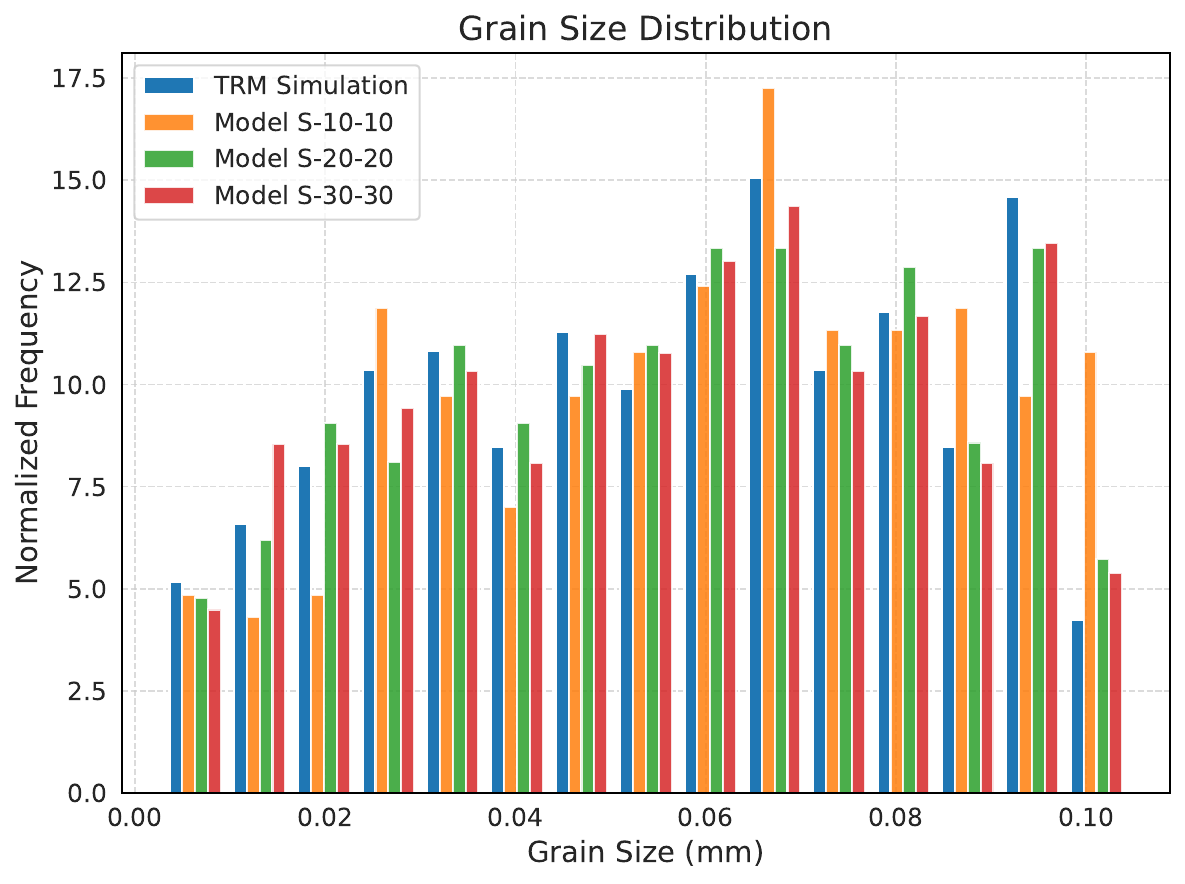}
        \caption{}
        \label{fig:freq_grain_dist}
    \end{subfigure}
    \hfill
    \begin{subfigure}[b]{0.49\textwidth}
        \centering
        \includegraphics[width=\textwidth]{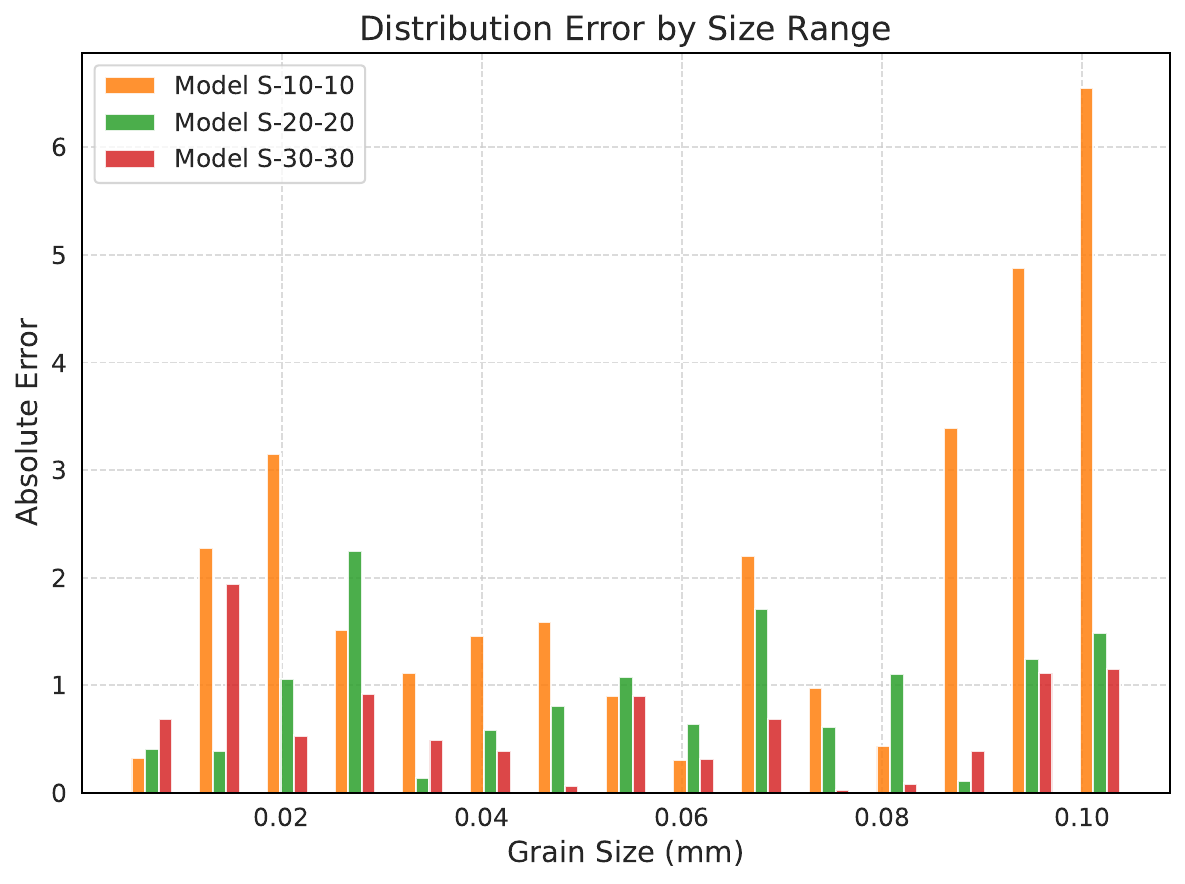}
        \caption{}
        \label{fig:freq_grain_dist_error}
    \end{subfigure}
    
    \vskip\baselineskip
    
    \begin{subfigure}[b]{0.49\textwidth}
        \centering
        \includegraphics[width=\textwidth]{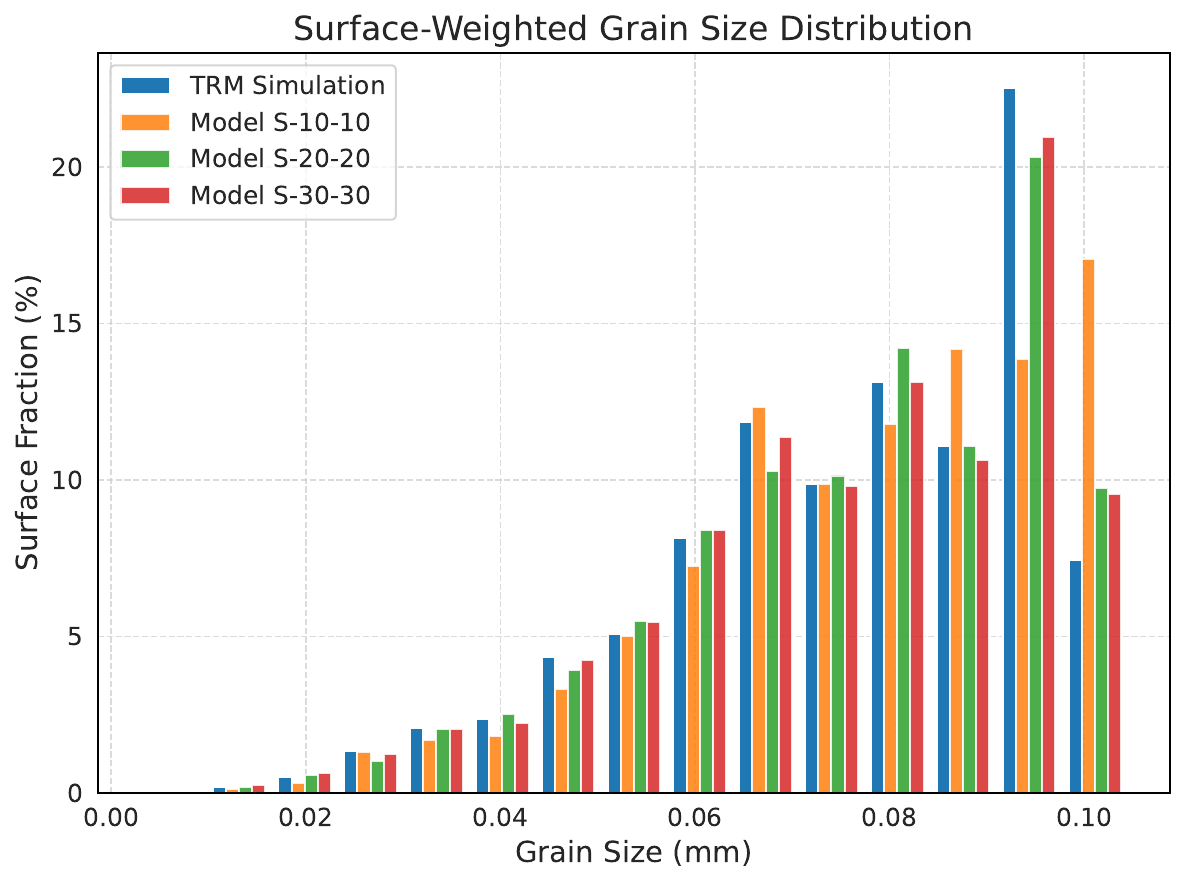}
        \caption{}
        \label{fig:sw_grain_dist}
    \end{subfigure}
    \hfill
    \begin{subfigure}[b]{0.49\textwidth}
        \centering
        \includegraphics[width=\textwidth]{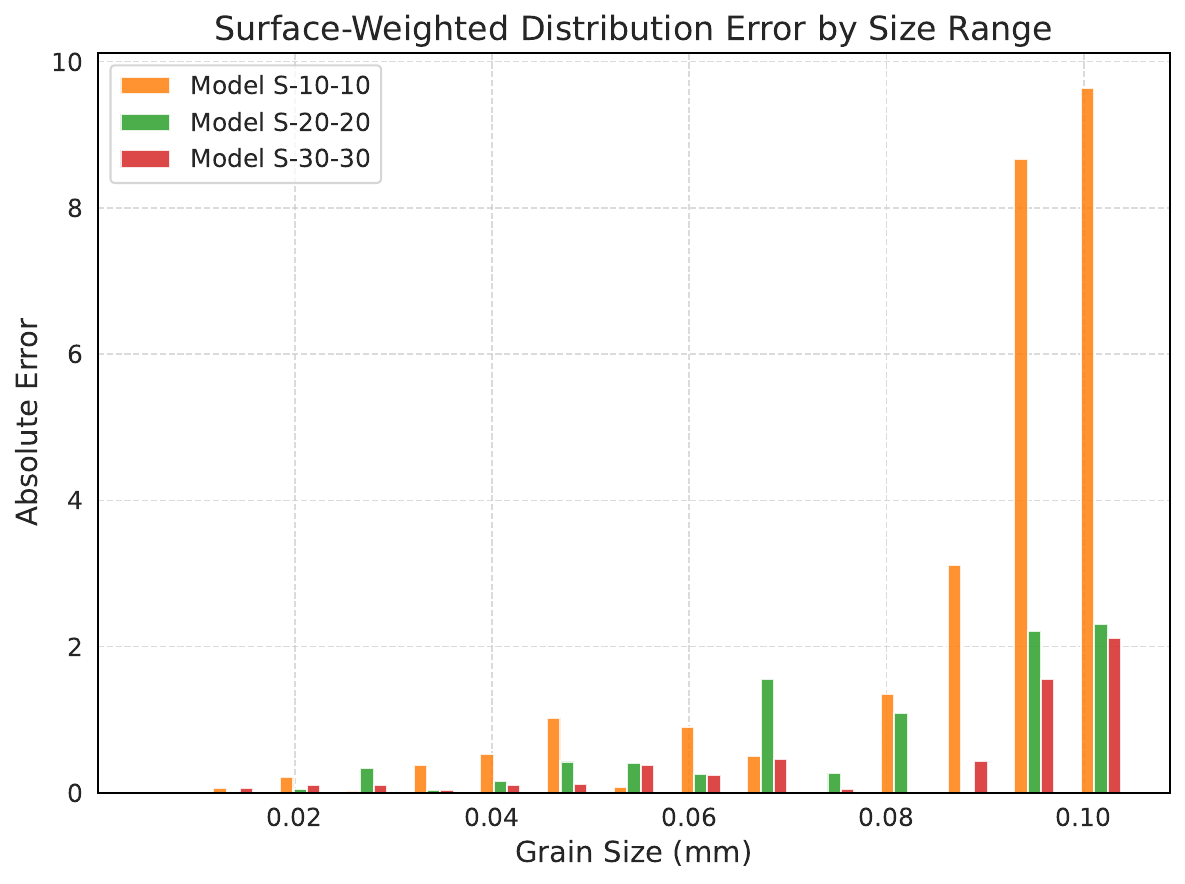}
        \caption{}
        \label{fig:sw_grain_dist_error}
    \end{subfigure}
    \caption{Grain size distribution analysis comparing Models S-10-10, S-20-20, and S-30-30 against TRM simulation. (a) Normalized frequency grain size distribution comparing model predictions with TRM simulation. (b) Distribution error by size range showing absolute errors in normalized frequency distribution for each model. (c) Surface-weighted ECR distribution showing the percentage of surface area contributed by different grain sizes for each model compared to TRM simulation. (d) Surface-weighted ECR distribution error by size range demonstrating absolute errors when grains are weighted by surface area.}
    \label{fig:grain_distribution}
\end{figure}

With regard to statistical metrics, Figure \ref{fig:grain_distribution} compares the grain size distributions predicted by each model with the conventional TRM simulation, both in terms of frequency and surface-weighted, along with error distributions by size range.

The ECR distribution weighted by normalized frequency in Figure \ref{fig:grain_distribution}(a) illustrated that all trained models follow the general trend of the ECR distribution of the conventional simulation method, peaking at around around \SI{0.06}{\milli\meter} to \SI{0.07}{\milli\meter}. The S-30-30 consistently demonstrates the closest approximation across all size ranges to the ground truth, as proven in Figure \ref{fig:grain_distribution}(b) with maximum absolute error values below 3. Conversely, the S-10-10 performs the worst, particularly for grain sizes above \SI{0.04}{\milli\meter}, where errors are consistently higher, peaking at \SI{0.10}{\milli\meter}.

The Surface-weighted ECR distribution presented in Figure \ref{fig:grain_distribution}(c) shifted the peak toward larger sizes (\SI{0.09}{\milli\meter}), which is expected due to the increased area contribution of larger grains. Despite all models closer approximation to the conventional TRM simulation, the S-10-10 still demonstrated the most significant deviations, especially at the largest bin, with an error exceeding \SI{9}{\percent} (Figure \ref{fig:grain_distribution}(d)). As anticipated, the S-30-30 continues to show the closest approximation to the ground truth across most size ranges as evidenced by Figure \ref{fig:grain_distribution}(c) and Figure \ref{fig:grain_distribution}(d). These findings match the statistical results listed in Table \ref{tab:model-performance}.

\begin{figure}[htbp]
    \centering
    \begin{subfigure}[b]{0.8\textwidth}
        \centering
        \includegraphics[width=\textwidth]{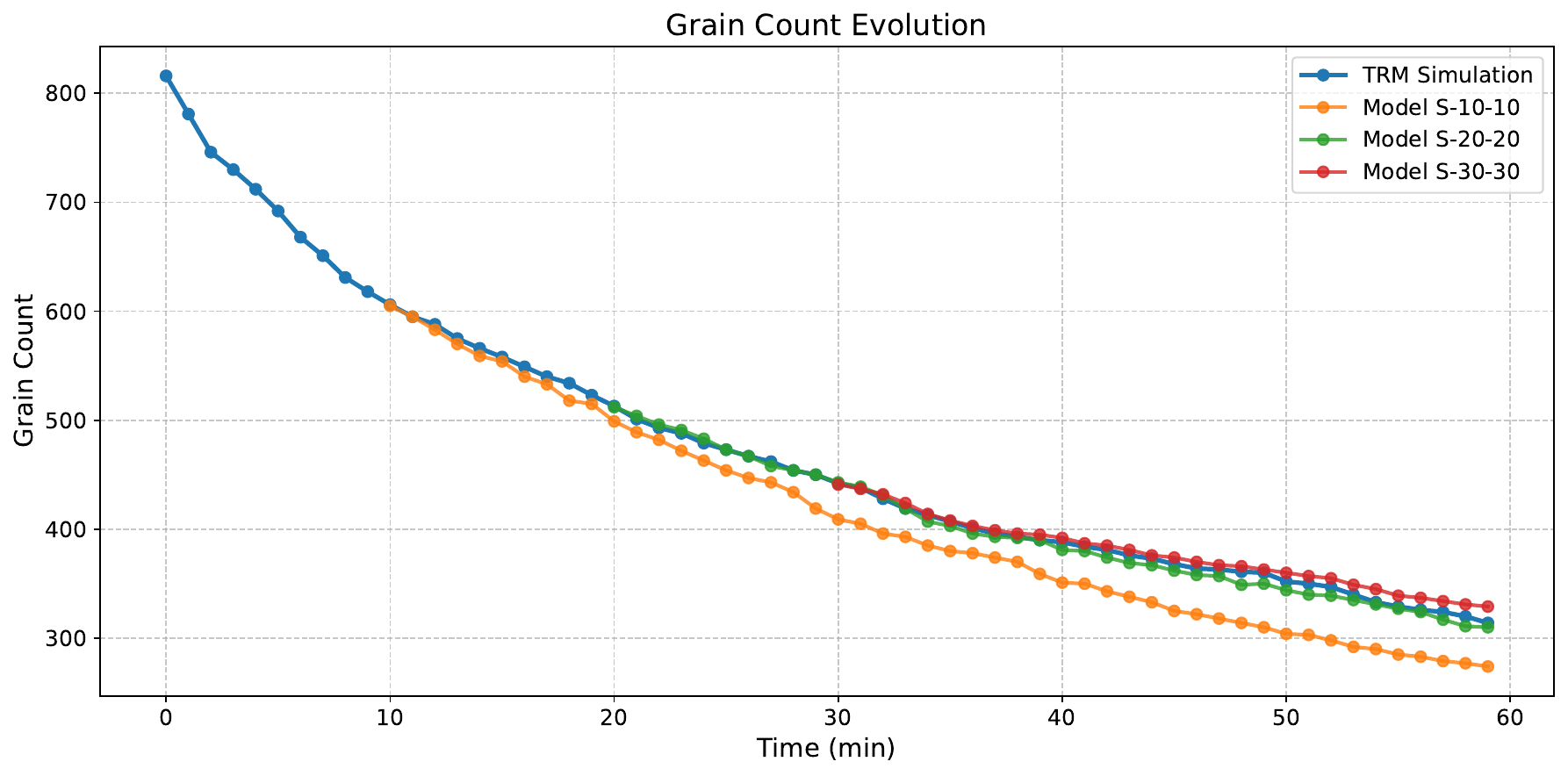}
        \caption{}
        \label{fig:grain_count_evo}
    \end{subfigure}
    
    \vskip\baselineskip
    
    \begin{subfigure}[b]{0.8\textwidth}
        \centering
        \includegraphics[width=\textwidth]{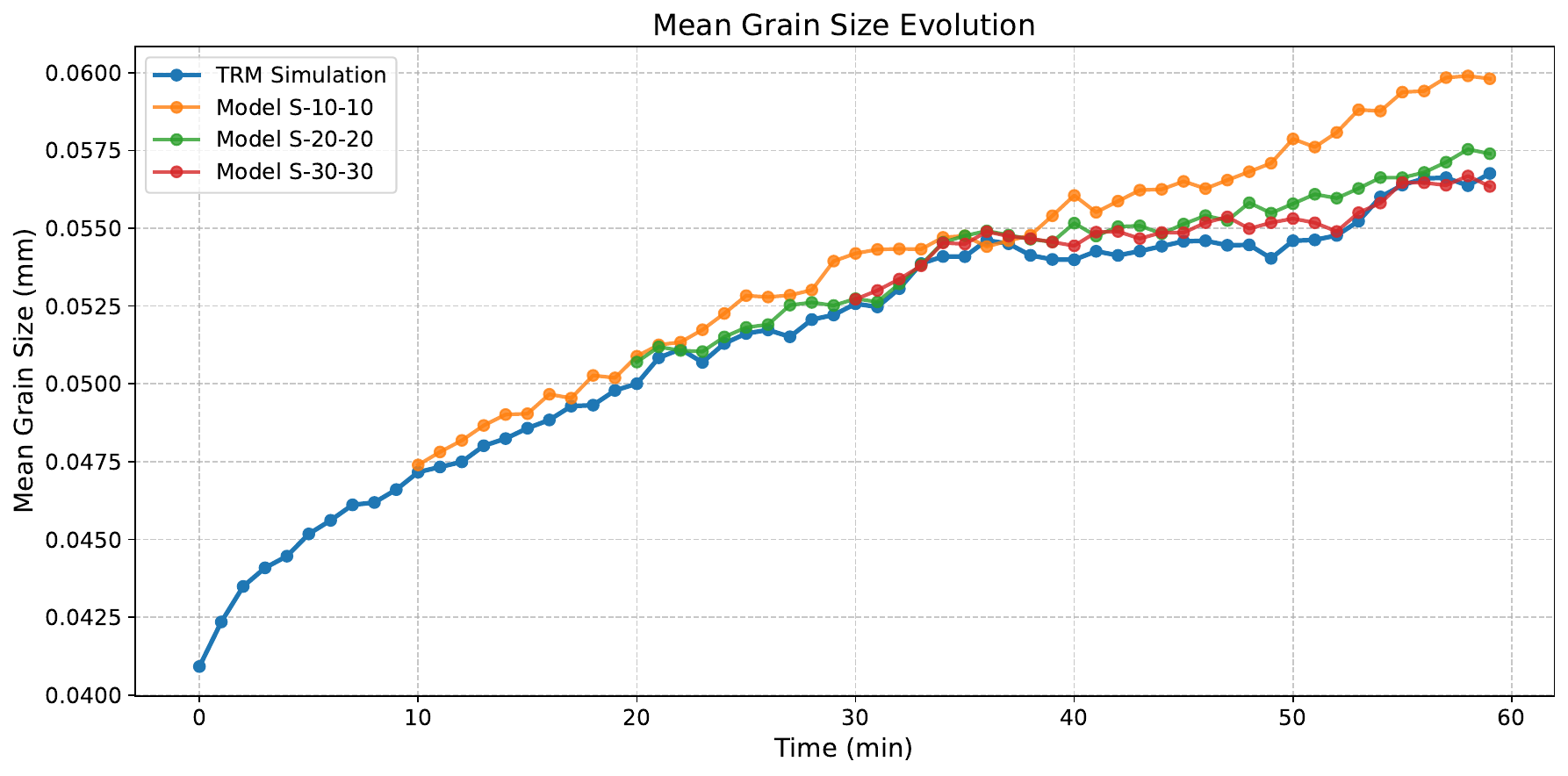}
        \caption{}
        \label{fig:mean_grain_size_evo}
    \end{subfigure}
    
    \caption{Temporal evolution of grains comparing TRM simulation with Models S-10-10, S-20-20, and S-30-30. (a) Grain count evolution. (b) $\overline{R}$ evolution.}
    \label{fig:grain_evo}
\end{figure}

To further validate the results, Figure \ref{fig:grain_evo} shows the temporal evolution of grain count and $\overline{R}$. The grain count evolution in Figure \ref{fig:grain_evo}(a) shows a consistent decrease in grain count over time. Initially, the conventional TRM simulation began with approximately 820 grains at initial state and gradually decreased to around 310-320 grains at $t=\SI{1}{\hour}$. The S-10-10 showed the most significant deviation with a steeper decline resulting in below 300 grains threshold by the end of the evolution. Meanwhile, the S-20-20 and the S-30-30 followed the trajectory of the conventional TRM simulation closely throughout the entire evolution, resulted in total number of grains above 300.

Furthermore, Figure \ref{fig:grain_evo}(b) exhibits an increase of $\overline{R}$ from \SI{41}{\micro\meter} to \SI{56}{\micro\meter} as the square root of time which is a nature of grain growth mechanism. The S-10-10 consistently overestimated mean grain size after $t=\SI{15}{\minute}$, peaking at approximately \SI{60}{\micro\meter}. In contrast, the S-20-20 and especially the S-30-30 maintained close and better performance in comparison to the S-10-10, with the S-30-30 providing remarkably accurate predictions, particularly between $t=\SI{40}{\minute}$ and $t=\SI{60}{\minute}$.

\begin{figure}[htbp]
    \begin{subfigure}{0.5\textwidth}
        \includegraphics[width=1\linewidth]{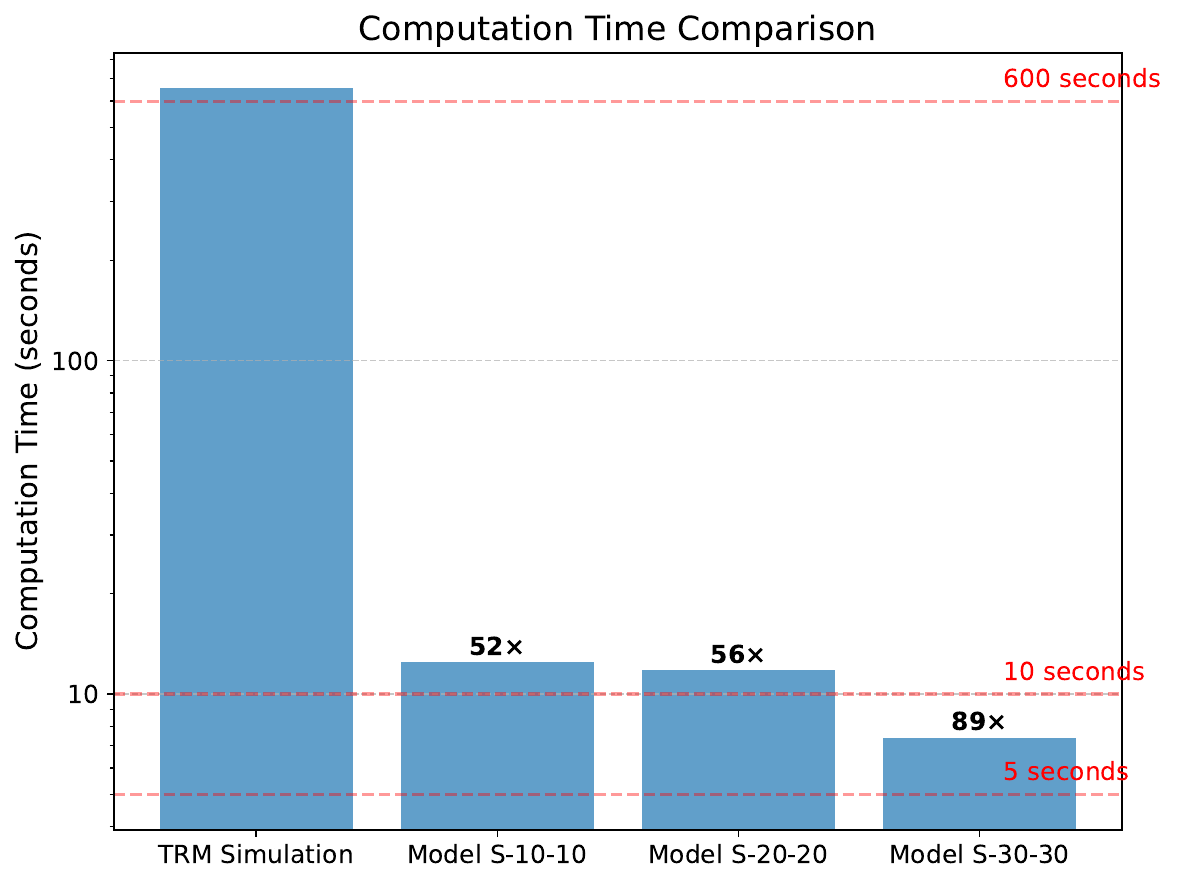}
        \caption{}
        \label{fig:computation_time_compared_to_convetional_method}
    \end{subfigure}
    \hfill
    \begin{subfigure}{0.5\textwidth}
        \includegraphics[width=1\linewidth]{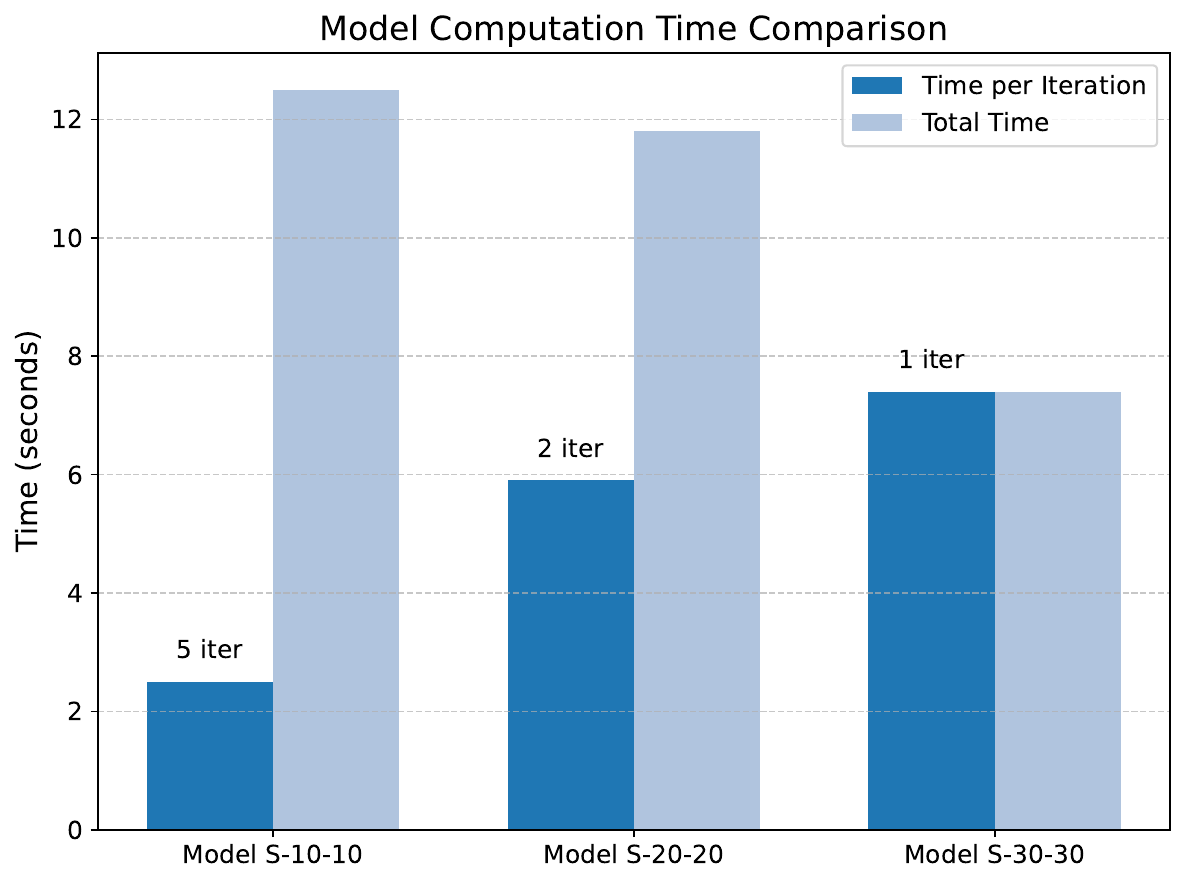}
        \caption{}
        \label{fig:computation_time_compared_ml_models}
    \end{subfigure}
    \caption{Computation time comparison between conventional method and ML models. (a) Logarithmic scale showing computation time of conventional method versus variants of ML models. (b) Linear scale comparison of ML models showing both per-iteration time and total computation time, with annotation of required iterations for each model configuration.}
    \label{fig:computation-time}
\end{figure}

A significant advantage of the ML approach over the conventional simulation methods becomes apparent when analyzing computational requirements, illustrated in Figure \ref{fig:computation-time}. To predict the complete evolution, the conventional TRM simulation took approximately \SI{600}{\second}, while the trained models reduced the computation time to under \SI{15}{\second}, achieving acceleration ranging from 52 and up to 89 times faster than the conventional TRM simulation. The S-10-10 showed the lowest time per iteration to predict the evolutions at \SI{2.5}{\second} per iteration yet required 5 iterations, resulting in a total computation time of approximately \SI{12.5}{\second} to predict the complete evolution. The S-20-20 demonstrated intermediate performance, requiring approximately \SI{6}{\second} per iteration, while the S-30-30 had the highest per-iteration computational cost at about \SI{7.5}{\second}, yielding total computation times of about \SI{12}{\second} and \SI{7.5}{\second}, respectively.

\section{Discussion}
\label{Discussion}
In this study, we evaluated the performance of three trained models based on ML approaches (S-10-10, S-20-20, S-30-30) for predicting grain growth evolution, in comparison to the conventional TRM simulation. In the following discussion, the results are interpreted in terms of: model performance; analysis of visual representations, grain size distribution, and temporal evolution analysis; computational efficiency; and concludes with the limitations and direction for future works.

Based on the results, the S-10-10 performed the worst across all evaluation metrics as evidenced by the model accuracy table, heatmap visualizations, and grain size distribution analysis. This lower performance was primarily due to the multiple prediction iterations required to reach the final state of the evolution at $t=\SI{1}{\hour}$. In each iteration, prediction errors from the previous step were carried forward and accumulated by additional uncertainties introduced during the next prediction. Over iterations, these accumulated errors led to significant deviations from the physical reality of the evolution in the final prediction. In contrast, the S-30-30 required only a single iteration to get to the final state, which greatly reduced the potential for error accumulation. This behavior highlights a common limitation in time-series prediction tasks which prediction quality tends to degrade over time, especially in models that rely on recursive prediction, where each future prediction is based on previously predicted outputs.

Despite the superior performance of the S-30-30, it is worth noting that the S-10-10 and the S-20-20 still achieved reasonably accurate results compared to the ground truth while required as less as first 10 minutes of evolution data to predict the final microstructure state at $t=\SI{1}{\hour}$. In real-world application or industrial context, this is a significant advantage as it can accelerate and make accurate long-term predictions using only a small portion of early-stage simulation data rather than waiting until the midpoint. While the S-30-30 produces the most accurate results, it does so at the cost of requiring a longer temporal window size.

Upon closer examination of the heatmaps, the grain boundaries junctions consistently exhibit higher error intensities, indicating that the models tends to predict the evolution at these regions less accurately than grain boundaries. This effect is particularly noticeable in the heatmap of the S-10-10 (Figure \ref{fig:heatmaps}(a)), where the deviations at junctions are more frequent and pronounced. In contrast, the S-30-30 demonstrated a cleaner and more accurate prediction that closely align with the ground truth at the regions. These variations in performance could be attributed to the distinct temporal window sizes used in each model configuration-the 30-minute input into the S-30-30 provided sufficient temporal context to model the rate and dynamics of grain evolution more accurately. On the other hand, the shorter 10-minutes window input into the S-10-10 could not fully capture the slower dynamics of the system, causing it to over-predict evolution rates, resulting in the junctions that evolved faster than they should. 

Nonetheless, even with limited temporal input, all models were able to capture critical events such as triple junction dynamics and grain elimination, which are fundamental to the grain growth mechanism. This suggests that while temporal context significantly influences prediction accuracy, the models have learned core features of the underlying physical mechanism of the grain growth.

Grain size distributions are critical for understanding material properties and processing behavior, as they directly influence mechanical characteristics of the material. The analyzed distributions shown in Figure \ref{fig:grain_distribution} allow for more robust evaluation of how each model captures and predicts the grain across different sizes.

The errors observed in grain size distributions can be attributed to artifacts presented in the microstructure images produced during the reconstruction by the network. These artifacts particularly affected small grains, often resulting in over-segmentation or under-segmentation during post-processing, which involved segmenting the predicted images into distinct grains for grain size distribution analysis.

The surface-weighted error pattern differs from the normalized frequency approach, with the S-30-30 demonstrating superior performance but with all models showing elevated error values in the 0.05-0.08 \SI{}{\milli\meter} grain size ranges. This suggests that accurate modeling of larger grains is particularly important when surface-dependent properties are the main area of interest. Future model improvements should focus on improving prediction accuracy in that size range to better predict surface-mediated material behavior.

In term of mean grain size evolution, the overestimation by the S-10-10 correlates directly with the accelerated reduction in grain count observed in Figure \ref{fig:grain_evo}(a), suggesting that the S-10-10 predicted more accelerated grain coarsening than what actually occurred. These findings completely align with the heatmap results discussed earlier and confirming that models with longer temporal context provide better accuracy and more reliable predictions of the underlying grain growth mechanism.

The pattern of computational requirements suggests that models requiring more input data naturally take more time to process the input sequence and infer the subsequent evolution compared to those models with shorter input sequence. Additionally, the per-iteration computation time did not increase exponentially as we progress from the S-10-10 to the S-30-30, despite the threefold increase in input data volume. This sublinear scaling could be attributed to efficient of the neural networks, where convolutions are optimized at extracting temporal features in parallel. The significant speedup offered by ML models over the conventional methods highlights how ML models can learn to approximate complex physical mechanisms without explicitly solving the underlying differential equations, creating surrogate models that capture essential dynamics at a fraction of the computational cost.

Despite the predictive capabilities demonstrated by all these models, it is important to note the trade-off between the size of the input temporal window and the performance. As evidenced, the S-10-10 performed worse than the other models due to error accumulation over multiple iterations and the limited temporal context available for predicting long-term evolution. In contrast, the S-30-30 benefited from a larger temporal window, suggesting that neural networks can model the underlying dynamics for the grain growth mechanism more effectively when given more contextual information. Nevertheless, despite its lower performance in evaluation metrics, the S-10-10 still produced visually reasonable representations of the grain boundary network and successfully captured key features of the  grain growth mechanism.

However, the presence of artifacts in tiny grains remains a limitation. When the model is given microstructure images consisting of densely packed grains, where the number of grain is high within a small domain, it struggles to accurately capture their evolution. This issue likely stems from challenges in reconstructing fine-grained features and resolving closely spaced boundaries during prediction. Addressing this limitation will be an important direction for future work, potentially through enhanced resolution strategies, improved loss functions that prioritize small-scale features.

Predicting long-term evolution from only a small fraction of the initial data remains a key area of interest. Ongoing research will focus on improving prediction accuracy while mitigating error accumulation over time, possibly through incorporating physical constraints into the learning process.

While these models have shown robust capability in predicting grain growth mechanism on simulated data, performance may degrade when applied to experimental data due to factors such as noise, grain boundary shape differences, and measurement inaccuracies. Future work will evaluate the generalization ability of these trained models on real experimental data, assessing their predictive consistency and accuracy.

Moreover, the models will be further tested under different processing conditions outside their training distribution, such as variations in temperature, initial microstructure configurations, or grain boundary mobility, to evaluate their adaptability and robustness.

\section{Conclusions}
\label{Conclusions}
In this study, we successfully accelerated the prediction of grain growth evolution using a ML approach by modeling the task as a spatio-temporal learning problem with a hybrid neural networks architecture. This architecture combines an Autoencoder, in which its encoder and decoder blocks compressed and reconstructed microstructure representation; and a ConvLSTM, which captured the temporal dependencies and learned the underlying patterns of the grain growth mechanism to predict future states.

Our results demonstrated that this approach enables accurate and efficient prediction of the grain growth evolution. Several key findings were also found from this study. First, prediction accuracy was influenced by both the size of the input temporal window and the number of prediction iterations. Models trained with smaller temporal windows size and required more prediction iterations yielded lower accuracy in prediction compared to those that incorporated larger temporal window sizes with fewer iterations. Second, long-term evolution predictions based on minimal input data resulted in accumulated errors over successive iterations, particularly in extended prediction scenarios. Third, heatmap analysis further revealed varying degrees of prediction differences across all models. The S-10-10 showed the most deviations from ground truth, while the S-20-20 and the S-30-30 models showed progressive improvements. These differences were especially present at multiple junctions, where the predictions tended to predict slightly accelerated kinetics compared to the true evolution. Lastly, a notable challenge was the presence of artifacts produced during the reconstruction by the Autoencoder, particularly affecting tiny grains. These artifacts resulted in over-segmentation or under-segmentation, thereby impacting downstream tasks such as grain size distribution analysis.

On a positive note, a compelling advantage of the ML approach was the substantial reduction in computation time. The trained models achieved remarkable speedups, ranging from 52 to 89 times faster compared to the conventional TRM simulation, reducing prediction times from minutes to mere seconds. Although models trained with smaller temporal window size required less computation per prediction iteration, this advantage was partially offset by the need for multiple iterations when predicting long-term evolution. Nonetheless, this significant computational speedup makes ML-based simulation of grain growth evolution far more practical, particularly for real-world applications where time constraints are critical. Notably, even with minimal input data, the models learned the underlying pattern of the grain growth mechanism and were able to predict long-term evolution effectively. Moreover, even the lowest-performing model still produced visually reasonable representations of the grain boundary network. 

Despite these promising results, improvements are necessary. Future work will put a focus on improving the precision of long-term predictions from minimal input data. Additionally, we aim to extend our approach to accommodate various grain size distribution pattern and representative volume element size. Lastly, improve the reconstruction quality to translate grain structures into more accurate grain size distributions and minimizing artifacts, particularly in regions with fine grains, will be essential to reduce over-segmentation and under-segmentation errors and improve the accuracy of derived metrics such as grain size distributions.

\section*{Data availability}
Data will be made available on request.

\section*{Acknowledgments}

The authors thank ArcelorMittal, Aperam, Aubert \&
Duval, CEA, Constellium, Framatome, and Safran companies and the
ANR for their financial support through the DIGIMU consortium and
RealIMotion ANR Industrial Chair (Grant No. ANR-22-CHIN-0003).

\bibliography{main}
\end{document}